\documentclass{moriond}

\def\NIMA{{\textit{Nucl. Instrum. Methods} A}}

\def\be{\begin{equation}}
\def\ee{\end{equation}}
\def\bea{\begin{eqnarray}}
\def\eea{\end{eqnarray}}

\usepackage{xspace}
\usepackage{amsmath, bm}
\newcommand{\labfig}[1]{\label{fig:#1}}
\newcommand{\reffig}[1]{\hyperref[fig:#1]{Figure}~\ref{fig:#1}\xspace}

\begin{document}
\vspace*{4cm}
\title{Improved performances of the NectarCAM, a Medium-Sized Telescope Camera for the Cherenkov Telescope Array}

\author{F. Bradascio\\for the NectarCAM collaboration and the CTA consortium}

\address{IRFU, CEA Paris-Saclay, Universit\'e de Paris-Saclay}
% \\
% 91191 Gif-sur-Yvette Cedex, France}

\maketitle\abstracts{
NectarCAM is a camera developed to detect Cherenkov light between 80~GeV and 30~TeV. It will equip the medium-sized telescopes (MST) of the Cherenkov Telescope Array Observatory (CTAO). The camera comprises 265 modules, covering a field of view of 8 degrees. Each module consists of 7 photomultiplier Tubes (PMTs) equipped with light guides and a front-end board performing the data capture. NectarCAM is based on the NECTAr chip, which combines a switch capacitor array sampling at 1GHz and a 12-bit Analog to Digital Converter (ADC). The NectarCAM camera is currently under integration in CEA Paris-Saclay (France). In this contribution, I focus on the ongoing performance tests for its characterization and calibration before deployment on the CTAO North site.}

\section{The NectarCAM camera}
% CTA will consist of two large arrays of imaging atmospheric-Cherenkov telescopes (IACTs), see Figure 1, using techniques to detect VHE γ-rays. These VHE γ-rays are observed via the Cherenkov light produced in particle cascades (showers) generated when they interact in the Earth’s atmosphere. 

The Cherenkov Telescope Array (CTA) is a planned  Imaging Atmospheric Cherenkov Telescope (IACT) which will be able to detect very high-energy (VHE) gamma-ray photons between few tens of GeV to above 100 TeV~\cite{CTAconcept}. VHE gamma rays are indirectly observed via the Cherenkov light produced in particle cascades generated when they interact in the Earth’s atmosphere. With its 10 times higher sensitivity, the CTA observatory will dramatically outperform present generation arrays of IACT.  It will consist of more than 60 telescopes, arranged in two array sites, one in the Northern Hemisphere (in La Palma, Spain) and one in the Southern Hemisphere (in Chile), allowing to have access to almost all the night sky.
% {\bf Comment JF actually $> 60$ telescopes (at least 13 in the North and 15+35 in the South}

% CTA will be the largest ground-based gamma-ray detection observatory in the world.
In order to span four decades of energy, three different Cherenkov telescopes will be used: Small-Sized Telescope (SST), Medium-Sized Telescope (MST) and Large-Sized Telescope (LST). %{\bf Comment JF Size $\rightarrow$ Sized} 
The MST telescopes at the CTA-North site will be equipped with the NectarCAM camera~\cite{nectarcam}. 

% The camera is currently under integration at the integration facility in CEA Paris-Saclay (France). The full camera will be equipped with 265 modules (1855 pixels).

% The NectarCAM camera has a modular design, with a basic element (``module") consisting of a focal plane module (FPM) and a front-end board (FEB). The focal plane module is composed by seven 1.5" R12992-100-05 Hamamatsu photomultiplier tubes (PMTs) associated high voltage and pre-amplification boards (HVPA), and equipped with Winston cones, light concentrators made with hollow cones. The Cherenkov light deposited in the camera is first detected in the focal plane, converted into electric signal by the PMTs and preamplified towards two gain channels. 
% The signal is then amplified a second time in the FEB, where it is split into three channels: low and high gain channels, and trigger channel (L0 trigger). 
% Each FEB is plugged to a back-plane board which produces the camera trigger decision (L1) processing the L0 signals of a 37 pixels region~\cite{trigger}. %(including a L0 signal from 6 neighboring FEBs)
% The L1 signal is then sent to the trigger interface board (TIB)~\cite{TIB}, which sends back a L1-Accept (L1A) signal to all the modules if the trigger is accepted. 
% % The events are time-stamped on the TiCkS module~\cite{ticks}, a dedicated White Rabbit based board. 
% % When the L1A signal is received by the FEBs, the full waveform for every pixel is read-out in the NECTAr chips, digitized and transferred over Ethernet to a camera server where the event is build.
 
% \bigskip
The core of the NectarCAM camera is the NECTAr chip~\cite{nectar0}, a switch capacitor array able to perform the sampling of the signal at 1 GHz, combined with a 12-bit analogue-to-digital convert (ADC). The NECTAr chip acts like a circular buffer, which holds the data until a camera trigger occurs. 
The readout of the current version of the NECTAr chip is responsible for the major fraction of the deadtime for the NectarCAM camera. For this reason, a new version of the NECTAr chip and of the front-end board (FEB) is currently under verification at the integration facility in CEA Paris-Saclay (France). The new version of the NECTAr chip can run in ping-pong mode, meaning that the analog memory is divided into two, with the new signals from the PMTs being written into one half while the other is being digitized. This reduces the deadtime of the camera by an order of magnitude, going from 7~$\mu$s to $0.7~\mu$s which translates to a $5.2\%$ and $0.5\%$ deadtime at 7~kHz, respectively.
By reducing the deadtime by an order of magnitude, it will be possible to operate the NectarCAM camera at a higher trigger rate and consequently to lower the energy threshold down to $\sim 50$~GeV if operating the camera at a trigger rate of 7~kHz~\cite{trigger}.
In these proceedings, we present three verification tests that have been performed in the darkroom at the integration facility in CEA Paris-Saclay (France) using two new FEB modules (version 6, hereafter FEBv6) with the upgraded NECTAr chips.

\begin{figure}[t]
    \centering
    \includegraphics[width=0.69\columnwidth]{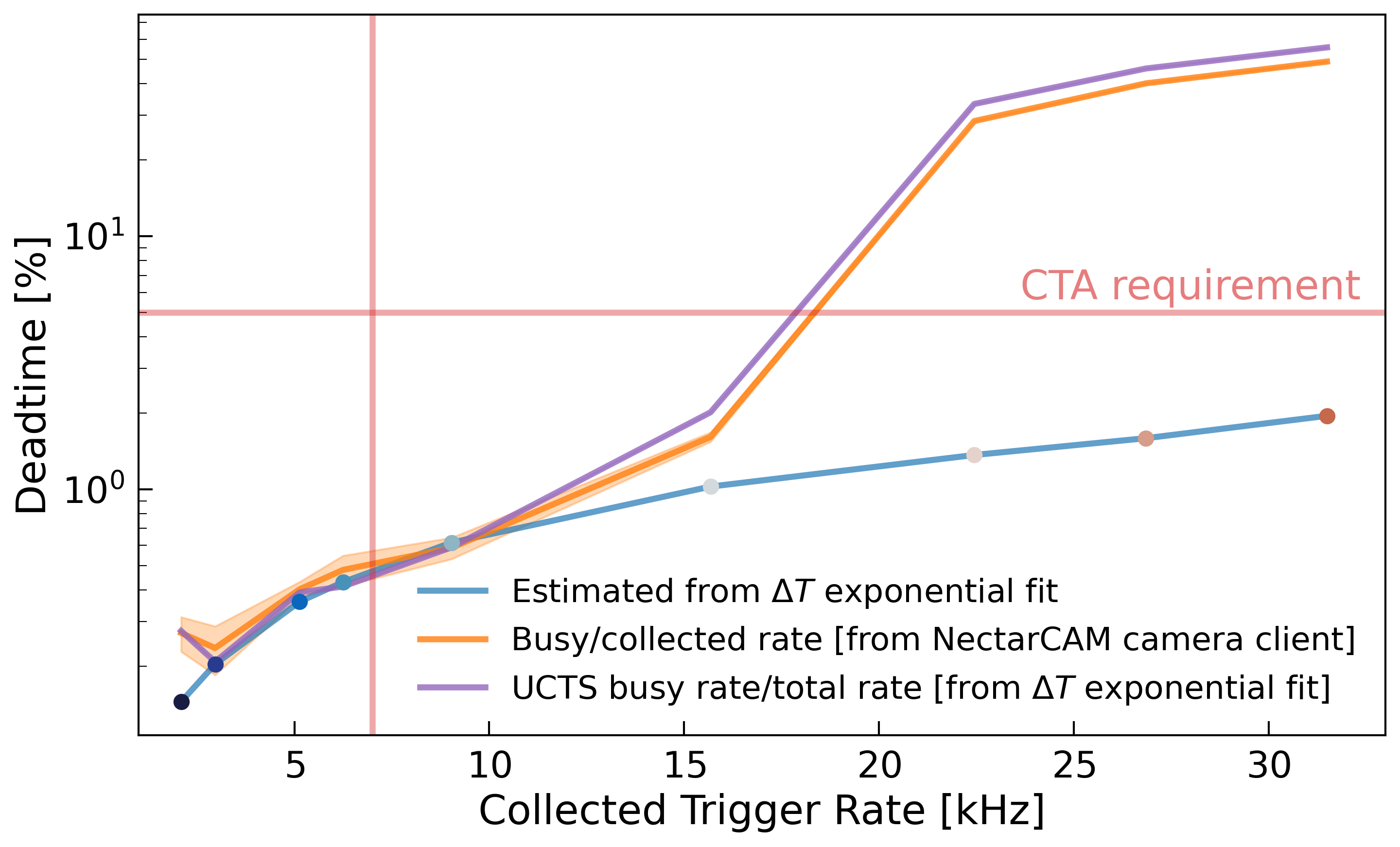}
    \caption{Deadtime fraction of two FEBv6. The deadtime estimated by the ratio between the busy trigger rate and the total trigger rate (orange and violet lines) and is compared with that obtained from an exponential fit (blue line). The CTA requirement of deadtime fraction at 7~kHz is shown by the red lines.}
    \labfig{deadtime}
\end{figure}

\section{The deadtime of FEBv6}

The reduction of the deadtime has been tested by measuring the smallest consecutive time $\Delta t$ between two events when the camera is illuminated with a random source. An exponential fit has been performed for different values of illumination intensity, resulting in an averaged deadtime of $\delta = 709.30 \pm 4.81$~ns. Since the NECTAr chip read-out is not anymore the limiting factor for the deadtime of the camera, this value is an upper limit to the true deadtime.
% For events occurring randomly according to a Poisson distribution, the probability that an event will occur at a time $[t, t+dt]$ with no events occurring between $t=0$ and time $t = \delta$ is ideally described by an exponential distribution, where $\delta$ represents the deadtime.
% In reality, some triggers could be lost when both the arrays of the ping-pong mode of the NECTAr chip are occupied. 
% However, we can still approximate the $\Delta t$ distribution with an exponential law. 
% An exponential fit has been performed for different values of illumination intensity, resulting in an averaged deadtime of $\delta = 709.30 \pm 4.81$~ns. 
By dividing this value by the trigger rate obtained from the exponential fit, the deadtime fraction is derived (see blue line in \reffig{deadtime}). This value is compared with the deadtime fraction obtained from the ratio between the busy trigger rate (i.e. the number of events which are triggered while the camera is busy writing data) and the total trigger rate (violet and orange lines). The two methods slightly differ from each other at high trigger rates, probably due to the fact that some busy triggers are lost or due to some overflows of the first-in-first-out(FIFO) queue system.
However, the two results agree up to a few tens of kHz, thus both fulfilling the CTA requirement of a deadtime $<5\%$ at 7~kHz.

\begin{figure}[t]
    \centering
    \includegraphics[width=0.7\columnwidth]{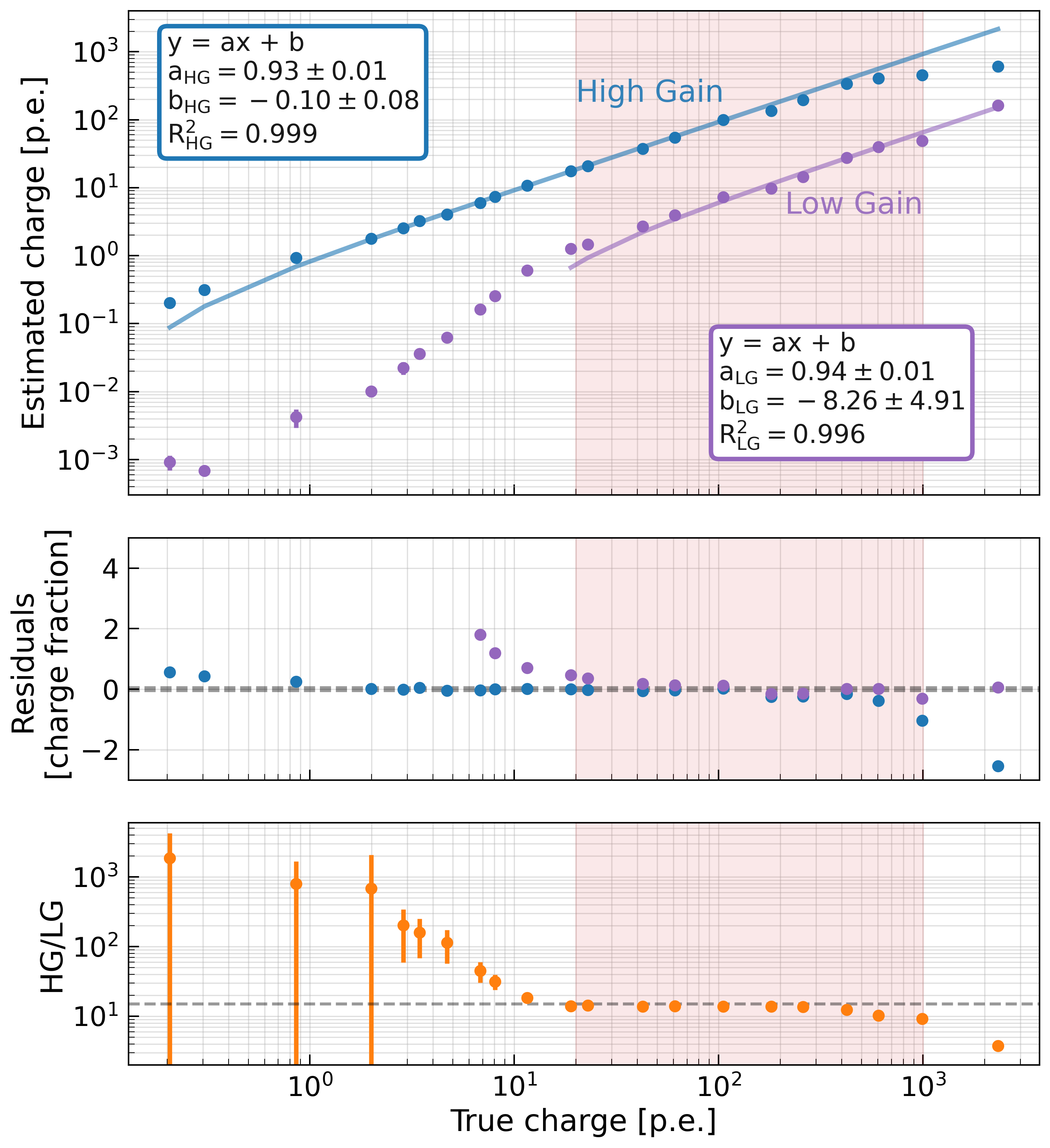}
\caption{Linearity of two FEBv6 modules. The top frame shows the measured charge as a function of the input pulse intensity for the high (blue points) and low (violet points) gain. The two linear fits with the corresponding parameters are shown. The fit residuals are displayed in the middle panel. The bottom panel shows the ratio between the two gains, and the red area indicates the overlapping range between the two gain channels of 20--1000~p.e.}
    \labfig{linearity}
\end{figure}

\section{The linearity of FEBv6}
The linearity describes the output distortion with the increase of the incident light intensity at a given gain. 
The linearity of the readout of the two FEBv6 has been measured by recording the charge deposited in every pixel when illuminated by a filtered LED source. A series of \textit{Edmund} filters~\footnote{\url{https://cdn.coverstand.com/30093/556052/1e10501890efd93e0b9a5b8a644dd99d07683283.pdf}} have been used to obtain an illumination in the range 0.1--3000~photoelectrons (p.e.).
% The deposited charge is obtained by integrating the PMT waveform in a readout window of 16~ns.  
The deposited charge is obtained by integrating the PMT waveform in a readout window of 16 ns, after subtraction of the baseline.

% When an event is triggered, the part of the PMT signal waveform sampled at 1 GHz which is inside the readout window of 16~ns is integrated, giving the “charge” deposited in each individual pixel.
The results are shown in \reffig{linearity}. The linearity it better than 5\% between 0.1 and 100~p.e. for the high gain, and between 100 and 2000~p.e. for the low gain. The overall dynamic range of this readout is therefore greater than 3 decades.
The high to low gain ratio remains constant at around 14 between 20 and 1000~p.e. (see bottom panel of \reffig{linearity}).

\section{The timing resolution of FEBv6}
The light's arrival time in each pixel is an important information that can be used to reduce the noise in shower images and improve the imaging cleaning and discrimination between Cherenkov photons and background. In this section, we estimate the timing resolution of the pixels of the two FEBv6. After illuminating the pixels with a uniform light created by 13 LEDs, the position of the maximum sample inside the 60~ns readout window (Time of Maximum, (TOM)) of each photon pulse has been measured for each pixel. 
The TOM is estimated from the waveform after subtracting the pedestal. Two methods have been used. The first one consists in identifying the position of the largest peak of the waveform using the function \texttt{signal.find\_peaks} from the \texttt{scipy} python package~\cite{scipy}. The latter performs a Gaussian fit of the largest peak of the waveform using as input the position of the peak obtained from the first method. 
The timing resolution of each pixel is given by the rms of the obtained TOM distributions. \reffig{timingres} shows the mean of the rms distribution over all 14 pixels as a function of the illumination charge. The points also include the 0.1~ns jitter of the LED source. For an incoming light of intensity above $20$ photons, both methods show that the time resolution is less than 1~ns, fulfilling the CTA requirement.
% As shown in \autoref{fig:waveform}, the two methods performs equally well. 
% The estimation on the light’s arrival time in all the pixels of a single camera image provides an additional and valuable information. This information may significantly help to reduce the noise in shower images and may improve the image cleaning, the reconstruction of the primary particle
% parameters or the gamma vs background discrimination. In this section we discuss the timing accuracy and systematic uncertainties in the NectarCAM camera. The time of maximum (TOM) of the Cherenkov pulse is estimated via a maximum likelihood method.

% the position of the maximum sample inside the readout window (called time of maximum, ToM)
\begin{figure}[h]
    \centering
    \includegraphics[width=0.69\columnwidth]{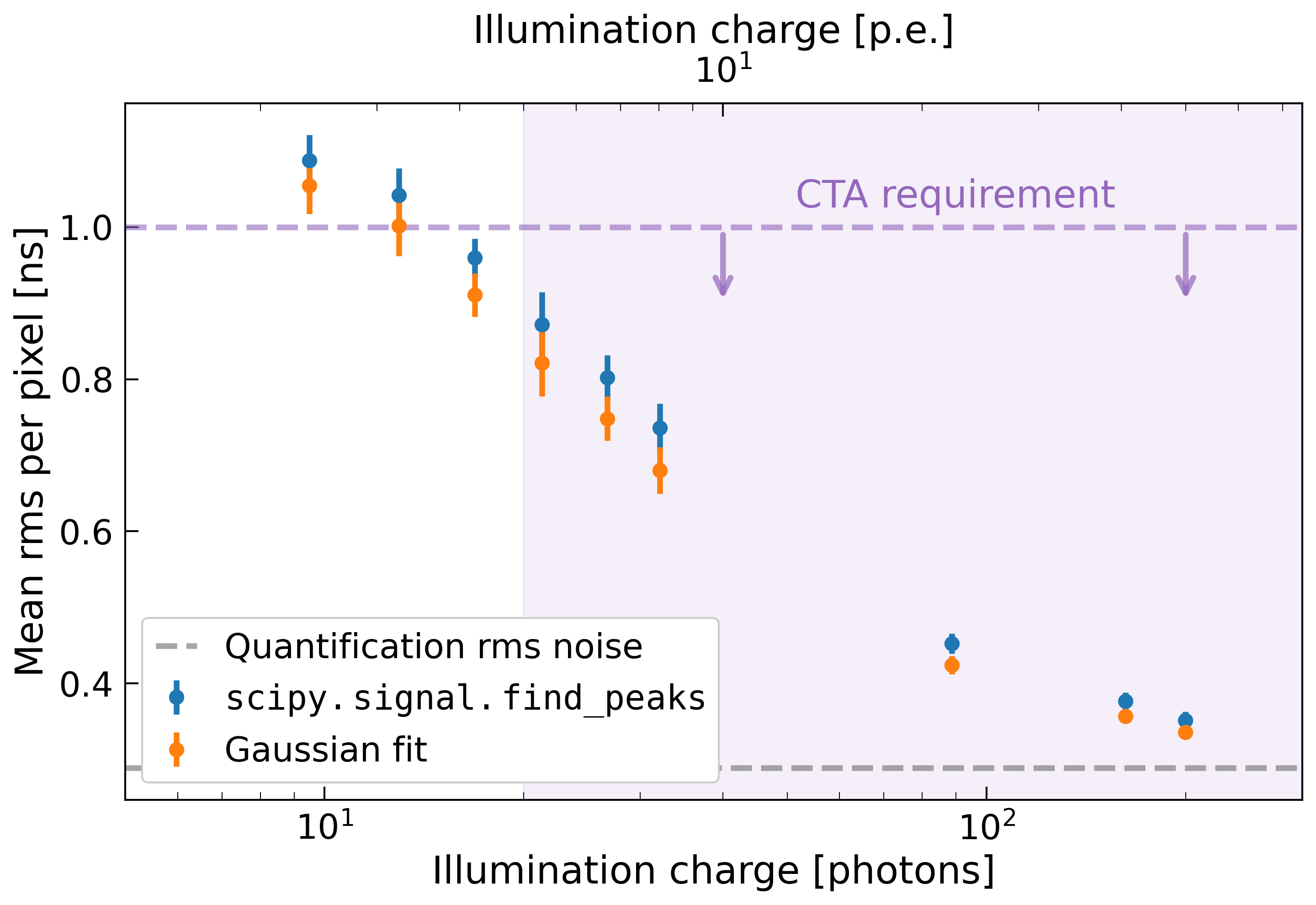} 
    \caption{Timing resolution per pixel (in ns) as a function of the charge of the illumination signal (in photons and p.e. on the bottom and top of the x-axis, respectively).  Both methods are shown (in blue and orange). The gray dashed line shows the quantification rms noise given by $\frac{1}{\sqrt{12}}$~ns. The dashed violet line shows the 1~ns requirement limit to be validated between 20[5] and 2000[500]~photons~[p.e.] (violet area).}
    \labfig{timingres}
\end{figure}

\section{Conclusions}
We presented the improved performances of two new FEBs to be used in the NectarCAM camera. The integration of an upgraded version of the NECTAr chip will reduce the deadtime of the camera by an order of magnitude with respect to the previous version. The new value is 0.7~$\mu$s, corresponding to a deadtime of $\sim0.5\%$ at 7~kHz. The linearity and timing resolution of the two new FEBv6 have also been tested. The former is better than 5\% over a range between 0.1 and 2000~p.e., the latter is $<1$~ns for a homogeneous illumination above 20~photons.

%%%%%%%%%%%%%%%%%%%%%%%%%%%%%%%%%%%%%%%%%%%
%%%%%%%%%%%%%%%%%%%%%%%%%%%%%%%%%%%%%%%%%%%
\section*{Acknowledgments}
This work was conducted in the context of the CTA Consortium. We gratefully acknowledge financial support from the agencies and organizations listed here: \url{https://www.cta-observatory.org/consortium_acknowledgments/}.
% \include{acknowledment}

%%%%%%%%%%%%%%%%%%%%%%%%%%%%%%%%%%%%%%%%%%%

\end{document}